\documentclass[12pt]{article}

\usepackage{amsmath,amssymb,graphicx,hyperref}
\usepackage[margin=1in]{geometry} 
\usepackage{authblk} 
\usepackage{float}
\usepackage{changepage}

\title{Stochastic Instabilities of the Diffusive Memristor}

\author[1]{Amir Akther\thanks{Corresponding author: A.Akther@lboro.ac.uk}}
\author[1]{Debi Pattnaik}
\author[1]{Yury Ushakov}
\author[1]{Pavel Borisov}
\author[1]{Sergey Savel'ev}
\author[1]{Alexander G. Balanov}
\affil[1]{Department of Physics, Loughborough University, Loughborough, LE11 3TU, United Kingdom}

\date{}

\begin{document}

\maketitle

\begin{abstract}
Recently created diffusive memristors have garnered significant research interest owing to their distinctive capability to generate a diverse array of spike dynamics which are similar in nature to those found in biological cells. This gives the memristor an opportunity to be used in a wide range of applications, specifically within neuromorphic systems. The diffusive memristor is known to produce regular, chaotic and stochastic behaviors which leads to interesting phenomena resulting from the interactions between the behavioral properties. The interactions along with the instabilities that lead to the unique spiking phenomena are not fully understood due to the complexities associated with examining the stochastic properties within the diffusive memristor. In this work, we analyze both the classical and the noise induced bifurcations that a set of stochastic differential equations, justified through a Fokker-Planck equation used to model the diffusive memristor, can produce. Finally, we replicate the results of the numerical stochastic threshold phenomena with experimentally measured spiking.
\end{abstract}

\section{Introduction}
Throughout the last ten years machine learning has emereged as one of the leading forces in modern scientific discovery and has permeated through a wide array of industries, including medical research~\cite{Medical}, banking~\cite{Finance} and transport~\cite{transport}. With this surge of modern applications along with the ever increasing volume of data required to train machine learning algorithms, we have slowly started to approach the constraints of the current computing architecture, that being a constant increase in the time to run newer computational tasks along with a growing amount of energy consumption to complete these tasks. With this continual increase of data and more complex algorithms being deployed we must find another approach to computing in order to sustain this trend. Fortunately, the field of neuromorphic computing shows great potential for overcoming such limitations. This computing method aims to replicate the ability to process information through techniques that are observable in living organisms which allow for exceptional pattern recognition at a remarkably low level of power consumption. The initial challenge is to create artificial neurons based on memristive technology that can mimic the high level of plasticity observed in real cells. These novel artificial neurons would then be used to produce a physical spiking network that is currently not feasible with this generation of software due to limits on resources. 

In 1971, Chua~\cite{Chua} hypothesized the existence of a new fourth fundamental circuit element which should exist out of the symmetries between the four circuit variables, specifically there should be a circuit element connecting charge and flux linkage. He named this newly hypothesized device the "memristor", this further got expanded when Chua and Kang~\cite{Kang} introduced a new set of dynamical systems with the term "memristive systems". The second landmark discovery in the field of memristive systems occurred in 2008 when HP labs~\cite{Hp} provided experimental evidence of the existence of the device. The memristor is a two-terminal component that exhibits a unique characteristic: its resistance varies in response to an internal variable. This resistance can undergo changes when a electrical current passes through the device, this resembles the observed resistance dynamics in biological cells and their associated synapses.

After the memristor was experimentally proven to exist, there has been a huge range of unique memristors which have been discovered. They either fall under the class of non-volatile or volatile, where the non-volatile memristors~\cite{NonVolatile} can remain within the same state of resistance after the applied voltage has been disconnected ~\cite{wang}. The secondary type of memristor, the volatile form, is in contrast to this, as when the external voltage is removed the memristor, after some short relaxation time, returns to it's initial resistive state~\cite{wang2017memristors}. By utilizing both forms of the memristor, the desired goal of producing an artificial neuron and synapse which is capable of being used to mimic biological neural networks may be feasible, and thus might be the optimal tools for achieving computing components that can be classified as neuromorphic devices~\cite{Wang2018}.

Following the discovery of the physical memristor, a wide array of distinct devices have been produced that would be categorized as volatile memristors~\cite{differentmechanism1, differentmechanism2, differentmechanism4}. The new device under investigation is known as the diffusive memristor, the electrical component switches resistance due to the diffusion of \(Ag\) clusters in \(SiO_{2}\) from one memristor terminal to the other, and this diffusion process results in a bridge being created from the left and right terminals~\cite{wang2017memristors}. A unique interaction between electrical, mechanical and thermal effects means that the diffusive memristor can display an assortment of interesting dynamical, stochastic and within certain systems chaotic dynamics(see, e.g., Refs~\cite{Yury,agnes}). This device therefore is both compelling for its mathematical properties but also due to the devices ability to replicate short and long term plasticity~\cite{wang2017memristors} often seen within biological cells.

There has been a surge in the production of "artificial neurons" which use an assortment of different memristors~\cite{spikingNb,spikingNb-stan,differentmemristor2,Pavel-Nb}. Some of these artificial neurons can exhibit consistent spike dynamics which makes them a promising component for oscillator based computing~\cite{Oscillatorbasedcomputing1,Oscillatorbasedcomputing2}.This is unfortunately not the case for the diffusive memristor due to its stochastic nature. However it is also known that biological neurons and synapses are inherently noisy~\cite{Biocircuit} and therefore some level of stochasticity within artificial neurons may be advantageous when attempting to mimic the processes that biological cells exhibit.

The primary goal for the memristive research community is to analyze the properties of interacting memristors, however within this paper we focus on a singular device. We investigate an artificial neuron which is based on the diffusive memristor in \cite{nat-com,Yury}. We perform a deterministic analysis of the spiking mechanisms, by neglecting thermal noise, which we analyze primarily through the lens of bifurcation theory. We also simulate the coupled stochastic differential equations to see what critical parameters the system requires to produce noise induced spiking, which allows us to gain an insight into how to control the spiking phenomena of the artificial neuron. Finally we analyze real world devices to see if our model qualitatively describes the same mechanism which leads to spiking.
\section{\label{sec:level2.1}Artificial neuron model}
The memristor under investigation in this paper is a two-terminal device in which \(Ag\) clusters (which may also be referred to as particles) are able to migrate between terminal; as a result of this, the resistance of the memristor is defined by the positions of the particles. A detailed description of the diffusive memristor ~\cite{wang2017memristors, Wang2018,memasym2017,nat-com} coupled to an electric circuit can be obtained using a model which takes into account the thermal effects along with the electron dynamics which influence the particles diffusion. A less complex version of the artificial neuron circuit model, where there is a singular \(Ag\) particle under consideration, the model is still accurate enough to capture the majority of the experimental results whilst allowing us to numerically investigate the obtained outcomes\cite{Yury}. The model is given:
\begin{equation} \label{eq:1}
\begin{cases}
\dfrac{dx}{dt} = -U'(x) + qV + \sqrt{\eta T}\xi(t), \\[8pt]
\dfrac{dT}{dt} =\dfrac{V^{2}}{\gamma R(x)}  - \kappa (T-T_{0}), \\[8pt]
{\tau_c}\dfrac{dV}{dt} = V_{ext} - \left(1+ \dfrac{R_{ext}}{R(x)} \right)V.
\end{cases}
\end{equation}

\begin{figure}[ht]
\includegraphics[width=0.85\textwidth]{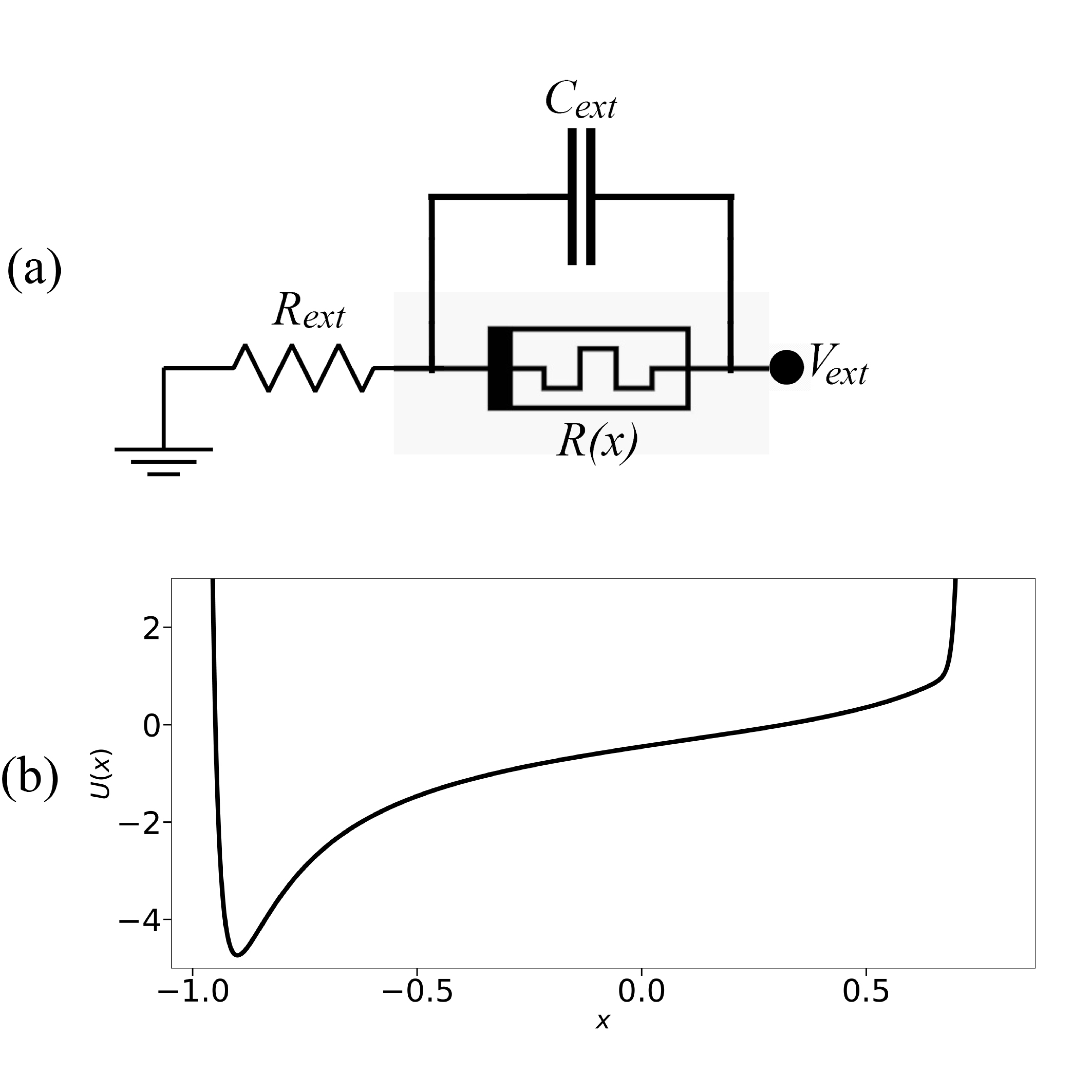}
\caption{\label{fig:potentialprofile2}a) Circuit diagram constructed through the use of Eq.~(\ref{eq:1}). b) \(U(x)\), representing the potential profile governing the dynamics of the particle using Eq.~(\ref{eq:potential}).}
\end{figure}

Here, \(x\) defines the particle position, the voltage between terminals is represented by \(V\) and T is the device temperature. We use \(q\) for the electrical charge of the particle, \(\xi(t)\) is a zero-mean \(\delta\)-correlated Gaussian noise, and the noise scaling parameter which influences the strength of the thermal noise which influences the particle is \(\eta\). The first equation in the above set governs the movement of the \(Ag\) cluster in the potential \(U(x)\) whilst being effected by the presence of thermal noise and an applied electric field . For simplicity we use \(U'(x)\) to represent \(dU/dx\).

The potential profile , \(U(x)\), displayed in Fig.~\ref{fig:potentialprofile2}(b) is modeled:

\begin{equation}\label{eq:potential}
U(x)= \dfrac{0.05}{(x+1)^{2}} - \dfrac{1}{x+1} - \dfrac{0.5}{x-1}  +(1.2x + 0.168)^{100}.
\end{equation}

The factors contributing to this potential come from (i) interactions between the \(Ag\) particle with inhomogeneities~\cite{ss-natcom} and the insulating matrix~\cite{ss-as-apl} and (ii) interfacial energy which favors the particle to get drawn towards immobile particles, forming filaments~\cite{ss-as1,ss-as2}, and the terminals of the memristor. The reason for the tilted geometry of \(U(x)\) (Fig.~\ref{fig:potentialprofile2}(b)) is motivated by experimental observations of hysteretic behavior within the $I-V$ curves~\cite{kumar2020third, kumar2017chaotic, messaris2020simplified}, which allows for the existence of both \(S-\) and \(N-\) type negative differential resistance. This is a critical property of the memristor which allows for the production of several different forms of current pulses~\cite{Ushakov-Akther}. 

The next equation within the model described in~(\ref{eq:1}) governs the temperature of the memristor. The equation captures the effects of the heat exchanging~\cite{Strukov} in the memristor with the Joule heat source, this exchange is effected by the internal resistance \(R(x)\) and with the heat capacitance coefficient \(\gamma\). Here, the heat transfer coefficient $k$ along with the temperature of the external bath \(T_{0}\), both determine the heat sink.

For the final equation which models voltage we utilize Kirchhoff's rule applied to the circuit displayed in Fig.~\ref{fig:potentialprofile2}(a). Here \(R_{ext}\), \(V_{ext}\) and \(\tau_c\) are used to define the applied resistance, voltage and the \(RC\)-time of the charging capacitor respectively. 

For the case of the system, the memristors internal resistance is dependent on the particles position. We make the assumption that the Ag ion is fluctuating in-between the platinum electrodes, and the electrons tunnel from each terminal through the single particle(this is reminiscent to the phenomena of electron shuttling(see Ref.~\cite{shuttle} for example); it follows that the resistance of the memristor fluctuates. Using this we can approximate the resistance of the particle as \(R(x)=cosh(x/\lambda)\) where \(\lambda\) defines the tunneling length.

The circuit displayed in Fig.~\ref{fig:potentialprofile2}(a) is described as an "artificial neuron"~\cite{Pavel-Nb} due to its ability to generate impulses that can be seen to resemble the spiking phenomena that can be seen in biological cells. Using Ohm's law, \(I=V/R(x)\), we can calculate the current within the device.

\section{\label{sec:level3.1}Spiking regimes}

Returning to the model described in Eq.~(\ref{eq:1}), previous works have shown that the memristor can produce two distinct forms of spiking~\cite{Yury}; the first type of spiking emerges due to cycles associated with charging and discharging whilst the secondary type is associated with thermal effects as the device can heat and cool which generates oscillations in the temperature.

The first form of spiking, by way of the charge-discharge cycle can be readily explained as such: the system starts at the high resistive state; when the \(V_{ext}\) is switched on, this causes the capacitor to charge due to the current flowing through it. When the applied voltage \(V_{ext}\) exceeds a critical parameter, the particle escapes the left potential well and rapidly reaches the center of the device, this is the point associated with the lowest value of resistance, \(R(x)\). This results in the capacitor to discharge, resulting in the reduction of the voltage. This reduction causes the resistance to increase which results in the particle returning to its original position and the capacitor begins to charge again. This charge-discharge cycle resembles the Pearson-Anson effect~\cite{Pearson}.

The systems other form of spiking is possible due to the interaction between the underlying deterministic system and the noise induced dynamics by thermal effects in the device. We are yet to understand the underlying mechanisms for both spiking regimes and we therefore utilize bifurcation analysis, both in the deterministic and stochastic sense to better grasp the mechanisms resulting in spiking.

\section{Results}
\subsection{\label{sec:level4.1}Bifurcation analysis}

\begin{figure}[H]
\centering
\includegraphics[width=15.5cm, height=6.5cm]{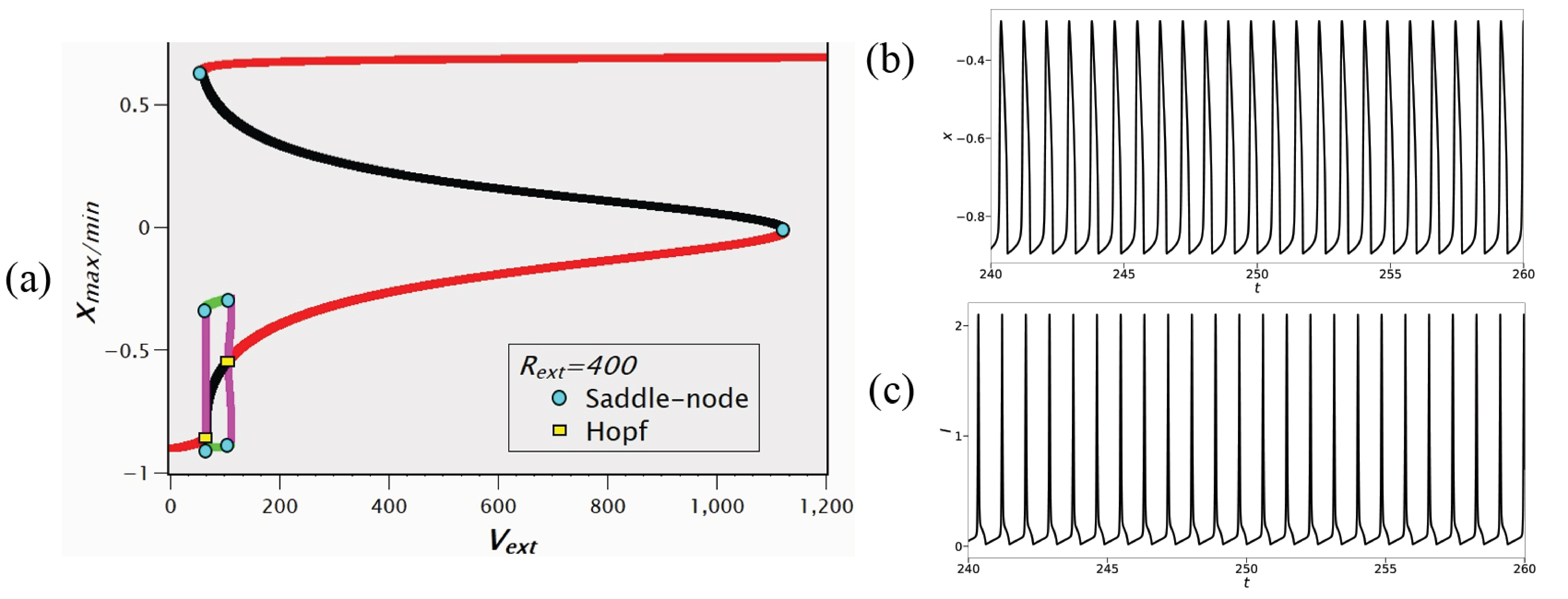}
\caption{\label{fig:Bifurcationdiagram} a) One-dimensional bifurcation diagram (at \(R_{ext} = 400\)). The stable and unstable fixed points are shown in red and black color respectively. Green illustrates the max/min values of the particle position,\(x\), on a stable orbit, pink is similarly defined for unstable orbits. b) Time series for $x(t)$ representing the spiking produced from the stable limit cycle. c) $I(t)$  representing the current oscillations which are produced from the deterministic charge-discharge cycle.}
\end{figure}

The model described in Eq.~(\ref{eq:1}) is capable of producing a range of interesting spiking dynamics by way of the interactions between deterministic and noise induced phenomena, however to uncover the purely deterministic methods that cause the onset of spike dynamics we must make modifications to the equation system. For this we must ignore the stochastic term in Eq.~(\ref{eq:1}), we do this by setting \(\gamma \rightarrow \infty\) and \(T_{0}\rightarrow 0\), thus reducing Eq.~(\ref{eq:1}) to a system of two equations only modeling particle position and voltage:
\begin{equation} \label{eq:2}
\begin{cases}
\dfrac{dx}{dt} = -U'(x) + qV , \\[8pt]
{\tau_c}\dfrac{dV}{dt} = V_{ext} - \left(1+ \dfrac{R_{ext}}{R(x)} \right)V.
\end{cases}
\end{equation}

The analysis of the mechanisms which give rise to spiking in model.~(\ref{eq:2}), is described in the bifurcation diagram shown in Fig.~\ref{fig:Bifurcationdiagram}(a). In the diagram the vertical axis represents the maximum/minimum of the particles position \(x\), the horizontal axis represents the applied voltage, \(V_{ext}\). For all further analysis we chose the following fixed parameters: $q=0.5$, $R_{ext}=400$, \(\tau_{c}\)=1 and \(\lambda=0.13\).

In the diagram, Fig.~\ref{fig:Bifurcationdiagram}(a), the bifurcation analysis uncovers three distinct chains of fixed points (denoted by red and black), the points illustrate the positions in which the artificial neuron remains stable. The fixed points of the artificial neuron are located around $x\approx-0.9$, $x\approx0$ and $x\approx0.68$, which represent the lower, middle and upper branches respectively. Given that the fixed points can coexist at the same external voltage \(V_{ext}\), it can be used to explain the multi-stability previously shown to exist in diffusive memristors along with the hysteresis in the transitions that can be observed in the current-voltage characteristics.

At the lower external voltage values, the system exhibits the only lower stable branch (red points) at \(x\approx-0.9\) which corresponds to the global minimum of the potential, \(U(x)\), shown in Fig.~\ref{fig:potentialprofile2}(b). The steady state remains constant the system undergoes an Andronov-Hopf~\cite{Bifurcationbook2, Bifurcationbook1} bifurcation occurs at \(V_{ext}\approx64.0\), where the system loses its stability for greater values of \(V_{ext}\). The Andronov-Hopf bifurcation produces an unstable limit cycle, whose maximum and minimum values of the particle position, \(x\), are given by the points in pink. The unstable orbit continues until a saddle-node bifurcation on a limit cycle occurs; this saddle-node changes the stability of the system resulting in the appearance of large stable periodic oscillations at \(V_{ext}\approx64.71\). These oscillations are the purely deterministic spiking which is caused by charge-discharge cycles within the circuit. The \(V_{ext}\) range which causes this starts at \(\approx64.71\) until \(\approx112.0\), where another saddle-node on a limit cycle occurs resulting in another unstable orbit which eventually ceases to exist as it merges to another Andronov-Hopf bifurcation at \(V_{ext}\approx106.2\). Therefore the underlying mechanism for the first spiking regime is the Andronov-Hopf and saddle-node bifurcation for periodic orbits. The aforementioned bifurcations have been previously shown to be associated with the charge-discharge spiking displayed by other memristive devices, e.g., \(NbO_{x}\) memristors analyzed in Ref.~\cite{Pavel-Nb}.

Looking at the bifurcation diagram it is clear that there are no other routes to spiking in the purely deterministic case, hence we must analyze the full equation system.~(\ref{eq:1}) to see what other underlying mechanism can result in another spiking regime.

\subsection{\label{sec:level5.1}Noise induced phenomena}
\begin{figure}[H]
\centering
\includegraphics[width=15.5cm, height=6.5cm]{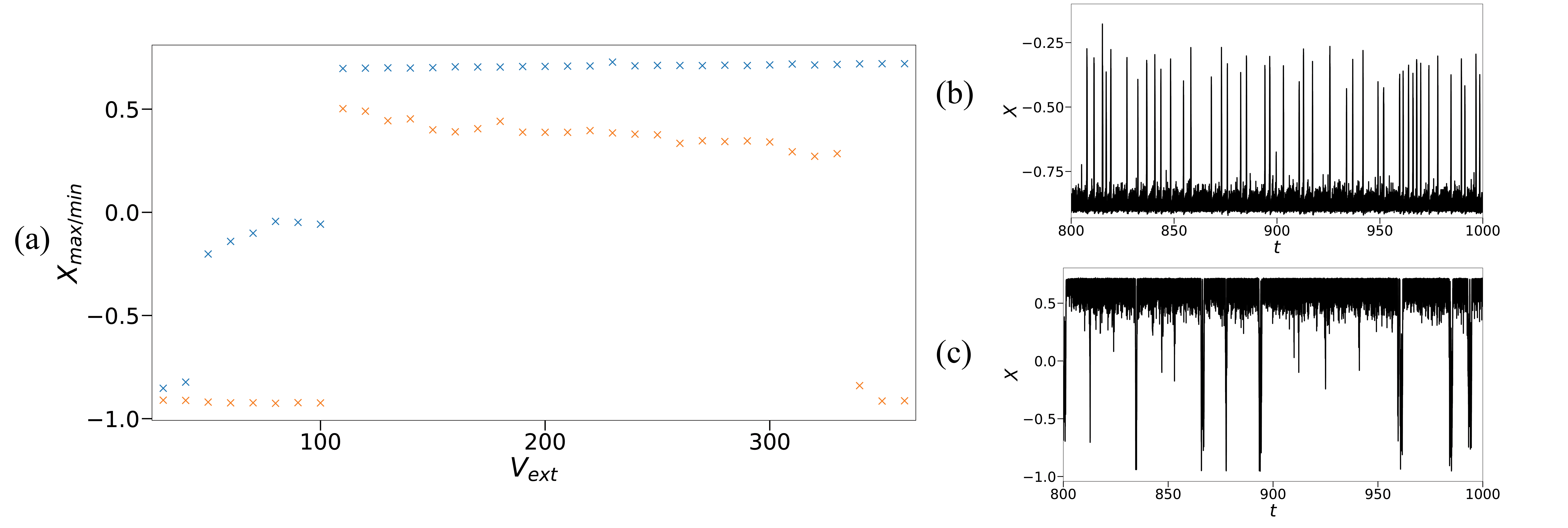}
\caption{\label{fig:OneDimensional} a) One-dimensional stochastic bifurcation diagram at \(R_{ext}=400\), blue crosses corresponding to the maximum particle position and orange crosses corresponding to the minimum position. The following additional parameters were chosen: $\gamma=5.555$ , $\kappa=0.2$ and $T_{0}=0$. b) Time series for $x(t)$ which corresponds to the stochastic version of the periodic charge-discharge cycles at $V_{ext}\approx50$. c) Time series for $x(t)$  displaying noise induced spiking at $V_{ext}\approx360$.}
\end{figure}

\begin{figure}[ht]
\includegraphics[width=0.86\textwidth,height=6.5cm]{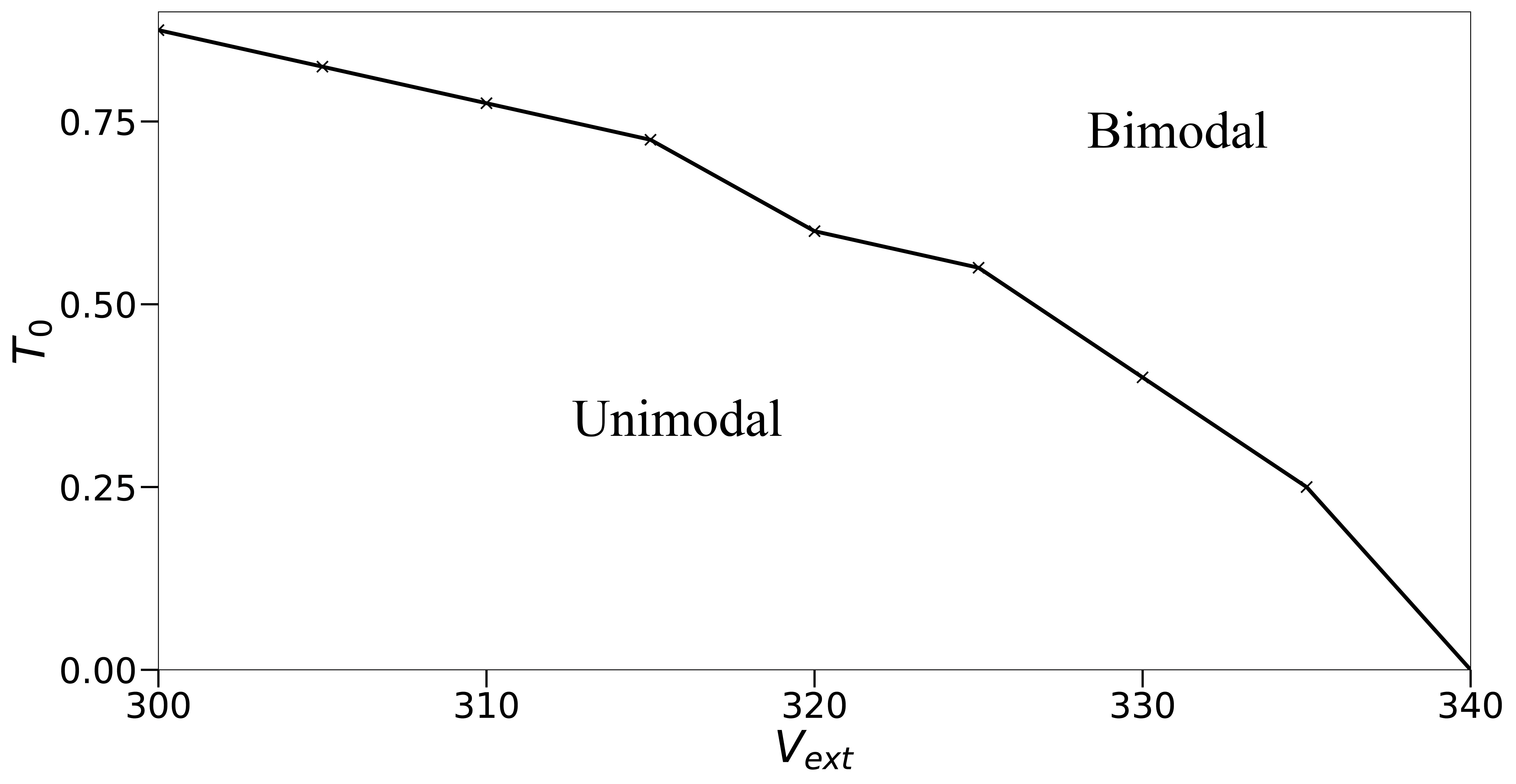}
\caption{\label{fig:TwoParameter} Two-parameter stochastic bifurcation diagram, the line of crosses corresponds to the points of the stochastic P-bifurcations.}
\end{figure}
We begin by analyzing the full equation system.~(\ref{eq:1}); whilst we removed temperature out of the system in the previous part we shall try to reproduce a Stochastic equivalent to the one-dimensional bifurcation diagram shown in Fig.~\ref{fig:Bifurcationdiagram}(a), the results are shown in Fig.~\ref{fig:OneDimensional}(a).

We start by fixing the particle position at \(x\approx0.68\), which exists in a region within the bifurcation diagram where the particle will be attracted to the bottom fixed point, \(x\approx-0.9\). We allow the system to run for a sufficiently long time period to eliminate transience and then calculate the maximum and minimum height that the particle achieves. The system is inherently stochastic and therefore there must always be a range of values of the position \(x\), we find the minimum and maximum positions (orange and black points respectively). We begin the analysis at \(V_{ext}\approx20\), the particle rapidly reaches the bottom fixed point, hence the first pair of points show simple fluctuations around the steady state, as seen in Fig.~\ref{fig:Bifurcationdiagram}. By slowly increasing the \(V_{ext}\), we see that at \(V_{ext}\approx50\), the max/min of the particle position, \(x\approx-0.2\)/\(x\approx-0.91\) respectively, equates to the max/min of the periodic orbit in the deterministic bifurcation diagram, the thermal noise within the system allows the particle to occasionally leap into the orbit, however given the external voltage is not at the critical parameter corresponding to the Andronov-Hopf bifurcation required to generate the self sustained oscillations the particle rarely spikes. With the increase in \(V_{ext}\) we see the bottom spiking regime take full form, the amplitude of the spikes increases due to the noise pushing the upper height of the deterministic oscillations out of its maximal limit cycle range. At \(V_{ext}\approx100\) we see that the particle gets stuck at the top fixed point; with the bottom fixed point we simply notice a max/min corresponding to noise fluctuations around the stable fixed point. The noise fluctuations around the fixed points increase in amplitude with an increase in \(V_{ext}\), however no noise induced stochastic bifurcation occurs until \(V_{ext}\approx340\), in which we see a drastic change in the max/min of the particle position. Within the deterministic bifurcation diagram we see this \(V_{ext}\) corresponds to the top fixed point, however the thermal noise pushes the particle out of this fixed point and results in the large spikes displayed in Fig.~\ref{fig:OneDimensional}(b). We therefore can conclude that the underlying mechanism that results in this drastic change of spike amplitude corresponds to the so called stochastic P-bifurcation~\cite{Arnold}, where the system undergoes a qualitative change in the stationary distribution of the Fokker-Planck equation. Whilst we don't directly solve the stationary solution we indirectly get an approximate stationary solution by simulating the stochastic differential equations for a sufficiently large time scale which should eliminate transients. 

We further analyzed the system by producing a two-parameter stochastic bifurcation diagram, as shown in Fig.~\ref{fig:TwoParameter}. The line of points correspond to the critical parameters, \(V_{ext}\) and \(T_{0}\) respectively, producing the P-bifurcation. These are the two most readily available to tune parameters for experimentalists, therefore it is crucial to uncover the parameter regions where we can control whether we observe spiking or not. 

\subsection{\label{sec:level6.1}Experimental evidence}
\begin{figure}[H]
\centering
\includegraphics[width=15.5cm, height=9cm]{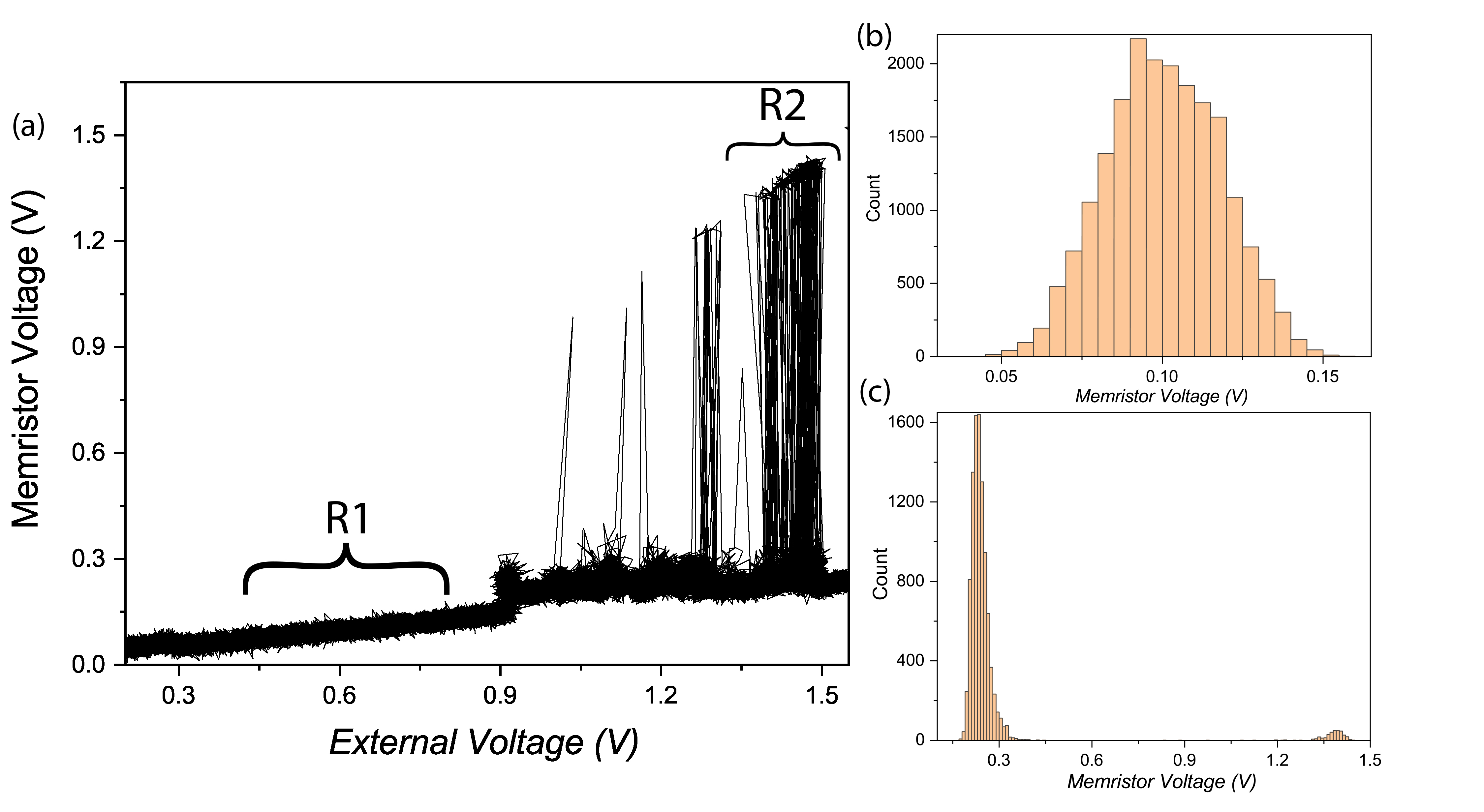}
\caption{\label{fig:Experiment} a) Experimentally measured memristor voltage against applied external voltage generating spikes past the voltage threshold. b) Histogram corresponding to the measured memristor voltage in \(R1\). c) Histogram corresponding to the measured memristor voltage in \(R2\).}
\end{figure}
To ensure that our model captures the correct noise induced phenomena we analyze real world devices to see if the same threshold phenomena occurs with respect to applied external voltage and whether the same probability distributions are realized.

The diffusive memristor samples were prepared using the magnetron sputtering technique. $50 nm$ Pt bottom electrode was deposited on a p-type $Si$ ($100$) wafer, followed by co-sputtering Ag and
$SiO_{2}$ to obtain a nominal thickness of $50 nm$. The layers were
deposited by RF sputtering for \(SiO_{x}\) and DC sputtering for $Ag$, respectively. Following this, a $30 nm$  Pt top electrode was deposited through shadow masks to obtain $100 \mu m$ devices. All of the deposition process was carried out in argon and at room temperature. Voltage spikes were performed using a Ricol Digital Function/Arbitrary Waveform Generator DG4162 and a Picoscope 3000 Series oscilloscope. A triangular waveform was applied and the corresponding device voltage was measured using the oscilloscope channels. 

In Fig.~\ref{fig:Experiment}(a) the measured voltage across the memristor is plotted against the increasingly applied external voltage. It is clear that when the applied voltage is less than a certain threshold, $V_{ext}\approx1.05$, there are no spikes, which corresponds to the range below $V_{ext}=340$ in Fig.~\ref{fig:OneDimensional}(a) where the device simply has noise fluctuating around some steady fixed point. This can be seen in Fig.\ref{fig:Experiment}(b) where the range, R1, of measured memristor voltage has been used to produce a histogram which resembles a Gaussian distribution; this closely replicates what we find in the numerical data. Past this threshold we begin to see the birth of some form of spiking phenomena. We take the measured memristor voltage in the range R2, which corresponds to a range of externally applied voltage values which can produce thermal spiking; this threshold corresponds to the point \(V_{ext}=340\) displayed in Fig.~\ref{fig:OneDimensional}(a). We use the range R2 to produce a histogram which shows a bimodal distribution, Fig.~\ref{fig:Experiment}(c). This change in distribution is the characteristic trait of the P-bifurcation and thus we can state that the experimental system exhibits the properties of such a bifurcation. There is a clear difference with regards to the amplitudes of the respective double peak, which is due to the small time period spent within each spike. This asymmetry was predicted by the numerical experiments.

\section{Discussion and conclusions}
In this paper we analyzed both the underlying deterministic and stochastic mechanisms for the onset of two different separate spiking regimes in the diffusive memristor. We uncovered that the main mechanism for the charge-discharge cycle is the Andronov-Hopf bifurcation and can be produced entirely within a deterministic setting. Our bifurcation analysis showed that there are no other deterministic paths to spiking, therefore we reintroduced thermal noise within the system. We uncovered that the mechanism which the system undergoes is the stochastic P-bifurcation in which the system goes from a unimodal distribution to a bimodal asymmetric distribution. We further validated our theoretical model by performing an analysis on real world devices which confirmed the same phenomena occurs experimentally. This work has given us an insight into how best to model the devices to control what deterministic and stochastic phenomena we want the diffusive memristor to exhibit, which is crucial for the further development of artificial neurons along with their application to real computing tasks.

\vspace{6pt} 



\section*{Author Contributions}
Conceptualization, A.A. and D.P.; methodology, A.A. and D.P.; software, A.A. and D.P.; validation, A.A., D.P. and S.S.; formal analysis, A.A.; investigation, A.A.; resources, D.P. and P.B.; data curation, A.A. and D.P.; writing---original draft preparation, A.A. and D.P.; writing---review and editing, A.A and D.P.; visualization, A.A. and Y.U; supervision, P.B., S.S and A.G.B; project administration, S.S.; funding acquisition, P.B., S.S. and A.G.B. All authors have read and agreed to the published version of the manuscript.

\section*{Funding}
This work was supported by The Engineering and Physical Sciences Research Council (EPSRC) (grant No EP/S032843/1).

\section*{Data Availability}
The data that support the findings of this study are available from the corresponding author upon reasonable request.

\section*{Conflicts of Interest}
The authors declare no conflict of interest. The funders had no role in the design of the study; in the collection, analyses, or interpretation of data; in the writing of the manuscript; or in the decision to publish the~results.

\bibliographystyle{plain}
\bibliography{aipsamp}

\end{document}